\documentclass[prl,amsmath,twocolumn,superscriptaddress,showpacs]{revtex4}
\usepackage{epsfig}

\usepackage{amssymb}
\usepackage{amsmath}
\usepackage{amsfonts}
\usepackage{bm}
\usepackage{graphicx}
\usepackage{verbatim}
\usepackage{color}

\newcommand{\bk}{{\bf k}}
\newcommand{\br}{{\bf r}}

\newcommand{\nn}{\nonumber}
\newcommand{\ud}{{\textrm{d}}}

\newcommand{\bq}{{\mathbf q}}

\newcommand{\bn}{{\mathbf n}}

\newcommand{\mt}{{\tilde{m}}}

\begin{document}
\title{Magnetic hard-axis ordering near ferromagnetic quantum criticality}

\author{F. Kr\"uger}
\affiliation{London Centre for Nanotechnology, University College London, Gordon St., London, WC1H 0AH, United Kingdom}
\affiliation{ISIS Facility, Rutherford Appleton Laboratory, Chilton, Didcot, Oxfordshire OX11 0QX, United Kingdom}
\author{C.~J. Pedder}
\author{A.~G. Green}
\affiliation{London Centre for Nanotechnology, University College London, Gordon St., London, WC1H 0AH, United Kingdom}

\date{\today}

\begin{abstract}
We investigate the interplay of quantum fluctuations and magnetic anisotropies in metallic ferromagnets. Our central result is that fluctuations close to a 
quantum critical point can drive the moments to point along a magnetic hard axis. 
As a proof of concept, we show this behavior explicitly for a generic two-band model with local Coulomb and Hund's interactions, and 
a spin-orbit-induced easy plane anisotropy. The phase diagram is calculated within the fermionic quantum order-by-disorder approach, 
which is based on a self-consistent free energy expansion around a magnetically ordered state with unspecified orientation. Quantum fluctuations 
render the transition of the easy-plane ferromagnet first-order below a tricritical point. At even lower temperatures, directionally dependent transverse
fluctuations dominate the magnetic anisotropy and the moments flip to lie along the magnetic hard axis.
We discuss our findings in the context of recent experiments that show this unusual ordering along the magnetic hard direction.
\end{abstract}


\maketitle

Fluctuations near to quantum critical points in itinerant ferromagnets (FMs) can have drastic and often surprising effects. They generically render 
\emph{a priori} continuous phase transitions first order at low temperatures \cite{Pfleiderer+01,Uemura+07,Otero+08,Taufourr+10,Yelland+11}. 
In many systems, quantum phase transitions are preempted by the formation of superconducting \cite{Saxena+00,Hassinger+08,Yelland+11}, 
modulated magnetic \cite{Brando+08}, or unusual spin-glass phases \cite{Westerkamp+09,Lausberg+12}. Other metallic FMs show 
an unexpected ordering along the magnetic hard axis \cite{Lausberg+13,Steppke+13}.

It is well understood that the first-order behavior arrises from the coupling of the magnetic order parameter 
to soft electronic particle-hole fluctuations, giving rise to non-analytic terms in the free energy \cite{Belitz+99,Chubukov+04,Belitz+05,Kirkpatrick+12}. 
Because of this interplay between low-energy quantum fluctuations, metallic FMs are very susceptible towards the formation of incommensurate 
magnetic \cite{Conduit+09}, spin nematic \cite{Chubukov+04} or modulated superconducting states \cite{Conduit+13}. This spatial modulation is 
associated with  deformations of the Fermi surface that enhance the phase space for low energy particle-hole fluctuations. The  
phase reconstruction can therefore be viewed as a \emph{fermionic} quantum order-by-disorder effect, which can be studied systematically by 
self-consistently calculating fluctuations around a whole class of possible broken-symmetry states \cite{Karahasanovic+12,Thomson+13,Pedder+13}.

The coupling to electronic quantum fluctuations can also have counter-intuitive effects upon the direction of the magnetic order parameter. 
A notable example is the partially ordered phase of the helimagnet MnSi, in which the spiral ordering vector rotates away from the lattice favored 
directions \cite{Pfleiderer+04,Kruger+12}. Similar effects are possible in homogenous itinerant FMs. This is suggested by recent 
experiments that show unusual ordering of magnetic moments along hard magnetic directions \cite{Lausberg+13,Steppke+13}.

The first example is YbRh$_2$Si$_2$, which is a prototypical system for studying antiferromagnetic quantum criticality, but exhibits strong FM fluctuations \cite{Ishida+02}. 
Interestingly, isoelectronic cobalt substitution for rhodium or hydrostatic pressure stabilize FM order along the hard axis at low temperatures \cite{Klingner+11,Lausberg+13}. 
This behavior has been interpreted as a combined effect of magnetic frustration and classical fluctuations \cite{Andrade+14}.  Quantum fluctuations 
close to a quantum critical point are potentially strong enough to drive a moment reorientation
even in the absence of frustration. This is supported by recent experiments on YbNi$_4$P$_2$ \cite{Steppke+13},
which show quantum critical fluctuations with a uniform magnetic susceptibility that is an order of magnitude larger along the easy direction. 
Remarkably, at very low temperatures -- well below the Kondo temperature -- this system displays a switch of magnetic response anisotropy 
similar to YbRh$_2$Si$_2$ and develops FM order along the hard direction. 

In this Letter, we demonstrate that fluctuation-driven moment reorientation from the easy towards the hard magnetic direction is a very generic phenomenon 
that is expected to occur in a large variety of itinerant FMs close to quantum criticality. The underlying mechanism has the same origin as the fluctuation-driven first-order 
behavior seen in practically all itinerant FMs at low temperatures, irrespective of microscopic details \cite{Kirkpatrick+12}. At mean-field level, spin-orbit (SO) coupling leads to 
magnetic anisotropy $F_\textrm{ani} \sim (m_\parallel^2-m_\perp^2)$ with $m_\parallel$ and $m_\perp$  the moment components along hard and easy  
directions, respectively. The coupling of the magnetic order parameter to soft electronic particle-hole fluctuations gives rise to a directional dependent, non-analytic free-energy 
contribution $\delta F_\textrm{ani} \sim  m^2 (m_\parallel^2-m_\perp^2)\ln (T/\mu)$, that competes with the mean-field anisotropy and dominates at sufficiently low temperatures.
The resulting switching of moments toward the magnetic hard axis opens up phase space for low energy, transverse spin fluctuations.

Our starting point is a generic two-band model with local intra-band Hubbard repulsion $U$, Hund's coupling $J_\textrm{H}$, and SO coupling $\lambda$ between the bands,  
%%%%%%%%%%%%%%%%%%%
\begin{eqnarray}
H & = & \sum_{\alpha=1,2}\sum_{\nu=\uparrow,\downarrow} \int_\bk \epsilon_\alpha(\bk)c^\dagger_{\bk\alpha\nu}c_{\bk\alpha\nu} \nn\\
& & + U\sum_{\br,\alpha} \hat{n}_{\br\alpha\uparrow} \hat{n}_{\br\alpha\downarrow} -J_\textrm{H}\sum_\br \hat{\mathbf{s}}_{\br1}\cdot \hat{\mathbf{s}}_{\br2}\nn\\
& & + \frac{\lambda}{2}\int_\bk \left( c^\dagger_{\bk,1\uparrow} c_{\bk,2\downarrow} - c^\dagger_{\bk,1\downarrow} c_{\bk,2\uparrow}+\textrm{h.c.}\right),
\label{Hamiltonian1}
\end{eqnarray}
%%%%%%%%%%%%%%%%%%%%%%%%%%%%%%%%%%%%%%
where the operators $c^\dagger_{\bk\alpha\nu}$ ($c_{\bk\alpha\nu}$) create (annihilate) an electron with momentum $\bk$ and spin $\nu$ in band 
$\alpha$, and
$\hat{n}_{\br\alpha\nu}=c^\dagger_{\bk\alpha\nu}c_{\bk\alpha\nu}$ and $\hat{\mathbf{s}}_{\br\alpha} = \frac12 \sum_{ \nu\nu^\prime}  c_{\br \alpha \nu}^\dagger \bm{\sigma}_{\nu \nu^\prime} c_{\br \alpha \nu^\prime}$ denote the occupation-number and electron-spin operators, respectively. Here $\bm{\sigma}_{\nu \nu^\prime}=(\sigma_x,\sigma_y,\sigma_z)_{\nu \nu^\prime}$, with $\sigma_i$ the standard Pauli matrices. For simplicity, we assume isotropic electron dispersions $\epsilon_1(\bk)=t k^2$ and $\epsilon_2(\bk)=\epsilon_1(\bk)+\Delta$ with a crystal field splitting $\Delta$. Tight-binding corrections to the dispersion do not lead to magnetic anisotropies or qualitatively change the  
phase diagram, as long as the system is far from instabilities due to nesting or van-Hove singularities \cite{Karahasanovic+12}.
The above SO term is of the standard form for an orbital multiplet that transforms as an $L=1$ angular momentum (e.g. the three $t_{2g}$ orbitals) \cite{Rozbicki+11}, 
projected onto the two low-lying bands \cite{Eremin+02}.

We first explain the {\it mean-field behavior of the free energy}.  For sufficiently large $U$, the ground state of the system is a FM, and 
the directions of the magnetizations $\bm{m}_1$ and $\bm{m}_2$ of the two bands are locked together by the Hund coupling $J_H$.
Without SO interaction ($\lambda = 0$), the mean-field free energy is independent of the direction of the total magnetization,
\begin{eqnarray}
F_\textrm{mf}^{(0)}& = & \frac{1}{1-\left(\frac{J_\textrm{H}}{2U}  \right)^2}\left[U(\mt_1^2+\mt_2^2) - J_\textrm{H}  \mt_1\mt_2\right]\nn\\
& & +\sum_\alpha \left( A_\alpha \mt_\alpha^2 + B_\alpha  \mt_\alpha^4 + C_\alpha  \mt_\alpha^6  \right).
\end{eqnarray}
Here we have introduced the weighted moments $\mt_\alpha = m_\alpha +\frac{J_\textrm{H}}{2U}m_{\overline{\alpha}}$, with 
$\overline{\alpha}=2$ for $\alpha=1$ and vice versa. The coefficients in this Landau expansion are given by
 $A_\alpha=U^2 \int_\bk n_F'[\epsilon_\alpha(\bk)]$, $B_\alpha=\frac{U^4}{12} \int_\bk n_F'''[\epsilon_\alpha(\bk)]$, and 
$C_\alpha=\frac{U^6}{360} \int_\bk n_F^{(5)}[\epsilon_\alpha(\bk)]$ with $n_F$ the Fermi function.
Due to the crystal field splitting, $\Delta$, between the two bands, we find two distinct magnetic transitions that are separated in temperature. 
Since we are interested in behavior near to the paramagnetic state, for large enough $\Delta$ we approximate the free energy, keeping only the 
classical contributions to the $\alpha =2$ band. In this case, $\tilde{m}_2 = \frac{J_H}{2U} \tilde{m}_1$.

The directional dependence arises from SO coupling, which we treat in second-order perturbation theory, assuming $\lambda/2\Delta\ll1$. 
This gives us an additional term
\begin{equation}
F_\textrm{mf}^{(\lambda)} = \frac{\lambda^2}{2 U^2}B_1\left(\mathcal{P}_\parallel-\mathcal{P}_\perp  \right) \mt_1\mt_2,
\label{Fmf}
\end{equation}
where we have defined the projectors onto the $y$-axis $\mathcal{P}_\parallel = n_y^2$ and the $xz$-plane $ \mathcal{P}_\perp = n_x^2+n_z^2=1-n_y^2$ 
in terms of a unit vector $\bn$ which parametrizes the magnetization direction $\bm{m}_\alpha = m_\alpha \bn$.
Since $B_1>0$ we obtain an easy-plane anisotropy at mean-field level.

To derive an expression for the {\it fluctuation contribution to the magnetic free energy}, we first express the action as a fermion coherent-state path integral over 
Grassmann fields $\Psi(\br,\tau)=[ \psi_\uparrow(\br,\tau),\psi_\downarrow(\br,\tau)]$. To decouple the interaction terms, we perform a Hubbard-Stratonovich 
transformation, introducing spin- and charge-fluctuation fields $\bm{\phi}_\alpha(\br,\tau)$ and $\rho_\alpha(\br,\tau)$. This leads to the action
\begin{eqnarray}
\mathcal{S} & = & \int_0^\beta\ud\tau\int\ud^3\br \left\{ U\sum_\alpha \left(\bm{\phi}^2_\alpha-\rho_\alpha^2\right)+J_\textrm{H} \bm{\phi}_1\cdot \bm{\phi}_2  \right.\nn\\
& & +\sum_\alpha \overline{\Psi}_\alpha\left[\partial_\tau-t\nabla^2-\mu_\alpha+U\left(\rho_\alpha-\bm{\varphi}_\alpha\cdot\bm{\sigma}\right)   \right]\Psi_\alpha\nn\\
& & \left.+i\frac{\lambda}{2}\left( \overline{\Psi}_1\sigma_y \Psi_2- \overline{\Psi}_2\sigma_y \Psi_1  \right)\right\},
\end{eqnarray}
where we have defined $\mu_1=\mu$, $\mu_2=\mu-\Delta$, and $\bm{\varphi}_\alpha:=\bm{\phi}_\alpha+\frac{J_\textrm{H}}{2U}\bm{\phi}_{\overline{\alpha}}$. 
The magnetizations of the bands are given by the zero-frequency components of the spin-fluctuation fields, $\bm{m}_\alpha=\bm{\phi}_\alpha(\br,\omega=0)$.

\begin{figure}[t!]
\begin{center}
\includegraphics[width= \linewidth]{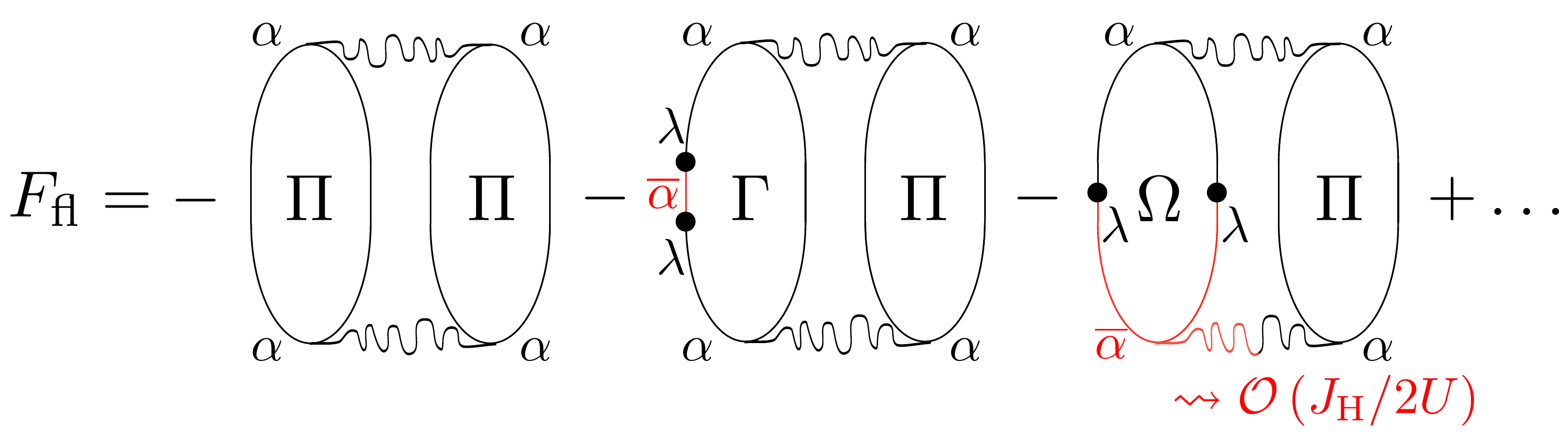}
\caption{(color online) Feynman diagrams for the fluctuation contributions to the free energy up to quadratic order in the SO  coupling $\lambda$. 
Solid lines indicate fermionic Green functions in the presence of FM order, wiggly lines denote propagators of the fluctuation fields.}
\label{fig.1}
\end{center}
\end{figure}

The key idea behind the fermionic quantum order-by-disorder approach is to include the magnetic order parameter in the free-fermion propagator and to 
self-consistently expand around the magnetically ordered, broken symmetry state. In this particular case, we expand around a magnetic state with an arbitrary 
magnetization direction, parametrized by the unit vector $\bn=(\sin\theta\cos\phi,\sin\theta\sin\phi,\cos\theta)$. The resulting free-fermion action is easily 
diagonalized by a rotation to new fermion fields $\eta_\alpha =\exp(i\theta\sigma_y/2)\exp(i\phi \sigma_z/2)\Psi_\alpha$, leading to 
$\mathcal{S}_0 = \sum_{\alpha\nu}\sum_\omega\int_\bk
 G^{-1}_{\alpha,\nu} \overline{\eta}_{\alpha,\nu} \eta_{\alpha,\nu}$ with Green's function
\begin{equation}
G_{\alpha\nu}(\bk,\omega) = [i\omega-( \epsilon_\alpha(\bk)-\nu U\mt_\alpha-\mu)]^{-1}.
\label{Green}
\end{equation}

In this new basis, the SO  terms explicitly depend upon the moment direction $\bn$. We perform the Gaussian integral over the fermion fields, 
keeping terms up to quadratic order in the finite-frequency fluctuation fields. We then integrate over the fluctuation fields to obtain
\begin{eqnarray}
\label{Ffl}
F_\textrm{fl} & = &   -\frac 12 U^2 T\sum_{\alpha,\tilde{\omega}}\sum_{\nu_1,\nu_2}\int_\bq \left\{  \Pi^{(\alpha)}_{\nu_1,\nu_2}\Pi^{(\alpha)}_{\overline{\nu}_1,\overline{\nu}_2}\right.\\
& & +\left.\lambda^2 \Pi^{(\alpha)}_{\nu_1,\nu_2}\left(  \mathcal{P}_\parallel \Gamma^{(\alpha)}_{\overline{\nu}_1,\overline{\nu}_2,\overline{\nu}_2}+\mathcal{P}_\perp \Gamma^{(\alpha)}_{\overline{\nu}_1,\overline{\nu}_2,\nu_2} \right) \right\}+ \dots,\nn
\end{eqnarray}
where, as before, we keep only terms up to quadratic order in $\lambda$. The corresponding Feynman diagrams are shown schematically in Fig.~\ref{fig.1}. 
Since the third diagram couples fluctuation fields from different bands it only contributes at higher-order in $J_H/2U$. We therefore neglect this term in the 
calculations below. In terms of the fermionic Green functions (\ref{Green}), the fermionic bubble diagrams are given by   
\begin{eqnarray}
\Pi^{(\alpha)}_{\nu_1\nu_2}(q) & = & T\sum_{\omega}\int_\bk G_{\alpha\nu_1}(k+q)G_{\alpha\nu_2}(k)\\
\Gamma^{(\alpha)}_{\nu_1\nu_2\nu_3}(q) & = & T\sum_{\omega}\int_\bk G_{\alpha\nu_1}(k+q)G^2_{\alpha\nu_2}(k)G_{\overline{\alpha},\nu_3}(k), 
\end{eqnarray}
where we have defined $k=(\bk,\omega)$ for brevity.

For $\lambda=0$, the bands decouple and $F_\textrm{fl}$  (\ref{Ffl}) reduces to the expression for the single band case \cite{Conduit+09,Karahasanovic+12}.  
After summation over Matsubara frequencies and re-summation of the leading divergencies in temperature to all orders in the magnetization we obtain \cite{Pedder+13} 
\begin{equation}
F_\textrm{fl}^{(0)} =\frac{c_1}{2} \frac{U^6}{\mu^5} \mt_1^4\ln\left( \frac{ \kappa U^2 \mt_1^2+T^2}{\mu^2} \right) -c_2\frac{U^4}{\mu^3}\mt_1^2,
\end{equation}
in agreement with the result of Belitz and Kirkpatrick \cite{Belitz+99}. We have neglected the fluctuation corrections for the second 
band, which is far from a magnetic instability, and defined $c_1=\frac{16}{3}\sqrt{2}(2\pi)^{-6}(\mu/t)^{9/2}$, $c_2=(1+\ln2)c_1$. Since fluctuations give rise 
to a $\ln T$-contribution to the the $m^4$ coefficient, the transitions turn first-order below a certain tri-critical temperature $T_c$. 

Although the fluctuation terms of order $\lambda^2$ appear to be much harder to calculate we can use a trick to rewrite 
$\Gamma^{(\alpha)}_{\nu_1\nu_2\nu_3}(q)$ in terms of a generalized derivative of $\Pi^{(\alpha)}_{\nu_1\nu_2}(q)$ \cite{derivativetrick}. From this, it is straightforward 
to see that the dominant directional-dependent fluctuation terms originate from \emph{transversal} spin fluctuations and are proportional to a derivative of 
$F_\textrm{fl}^{(0)}$,
\begin{equation}
F_\textrm{fl}^{(\lambda)} = \frac{\lambda^2}{8\Delta^2}\left(\mathcal{P}_\parallel-\mathcal{P}_\perp  \right) \mt_2 \partial_{\mt_1} F_\textrm{fl}^{(0)}.
\label{Ffl2}
\end{equation}

Since the magnetization of the system (global minimum of $F_\textrm{mf}+F_\textrm{fl}$) is always smaller than the value at which the function 
$F_\textrm{fl}^{(0)}$ has its minimum, the derivative in Eq.~(\ref{Ffl2}) is negative. Hence the directionally-dependent fluctuation terms compete with 
the mean-field anisotropy and could potentially stabilize FM order along the hard axis at sufficiently low temperatures.

\begin{figure}[t!]
\begin{center}
\includegraphics[width= \linewidth]{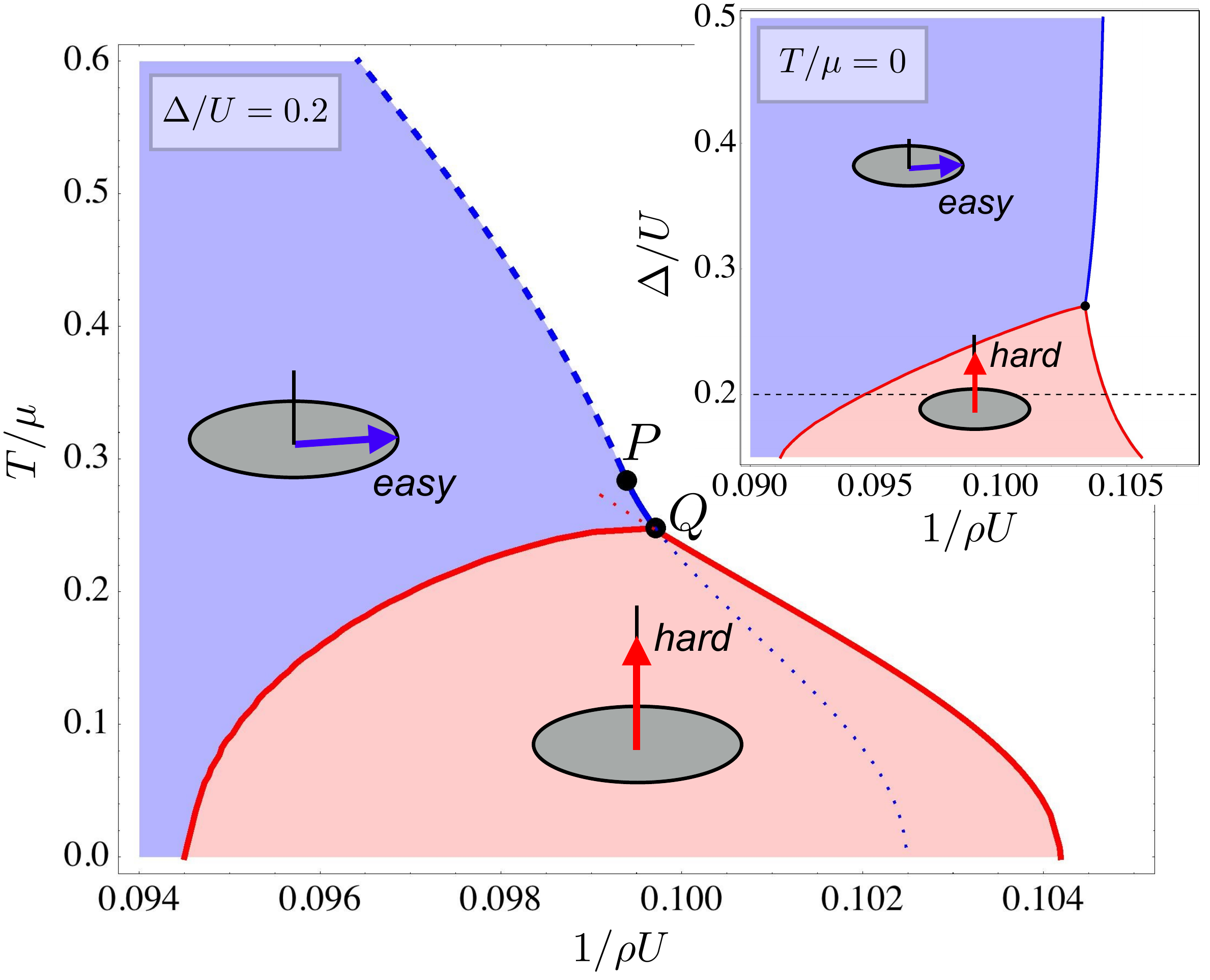}
\caption{(color online). Phase diagram as a function of temperature $T/\mu$ and inverse interaction strength $1/\rho U$ ($\rho$ is proportional to the density of states  
of band $\alpha=1$ 
at the Fermi level) for  $J_\textrm{H}/2U =0.3$, $\lambda/U=0.2$, and $\Delta/U=0.2$. At the tricritical point $P$ the nature of the phase transition
of the easy axis FM (blue region) changes from second to first-order, indicated by dashed and solid lines, respectively. In the red region below the point 
$Q$ fluctuations stabilize FM order along the hard axis. This region is enclosed by first-order transitions. 
The inset shows the $T=0$ phase diagram for the same parameters as a function of inverse interaction strength $1/\rho U$ and crystal field splitting $\Delta/U$.}
\label{fig.2}
\end{center}
\end{figure}

We calculate the {\it phase diagram} by minimizing the free energy $F=F_\textrm{mf}^{(0)}+F_\textrm{mf}^{(\lambda)}+F_\textrm{fl}^{(0)}+F_\textrm{fl}^{(\lambda)}$.
For small SO coupling $\lambda$ we can approximate $\tilde{m}_2\approx \frac{J_H}{2U}\tilde{m}_1$. The free energy then becomes a function 
of $\tilde{m}_1$, which to leading order in $J_H$ is the total magnetization of the system. In Fig.~\ref{fig.2} the phase diagram is shown as a function of temperature 
and inverse interaction strength.  Due to the non-analytic fluctuation correction $F_\textrm{fl}^{(0)}$, the transition of the easy-plane FM turns 
first-order at temperatures below the tricritical point $P$. At the point $Q$, located at a lower temperature,  we find a crossing between the first-order lines 
calculated from the free energies $F_\perp$ and $F_\parallel$ for moments in the easy plane and along the hard direction, respectively. Consequently, there is a region
below the point $Q$ where fluctuations stabilize magnetic order along the hard axis. As one should expect, the transition at which 
the moments flip is first-order. 

\begin{figure}[t!]
\begin{center}
\includegraphics[width= \linewidth]{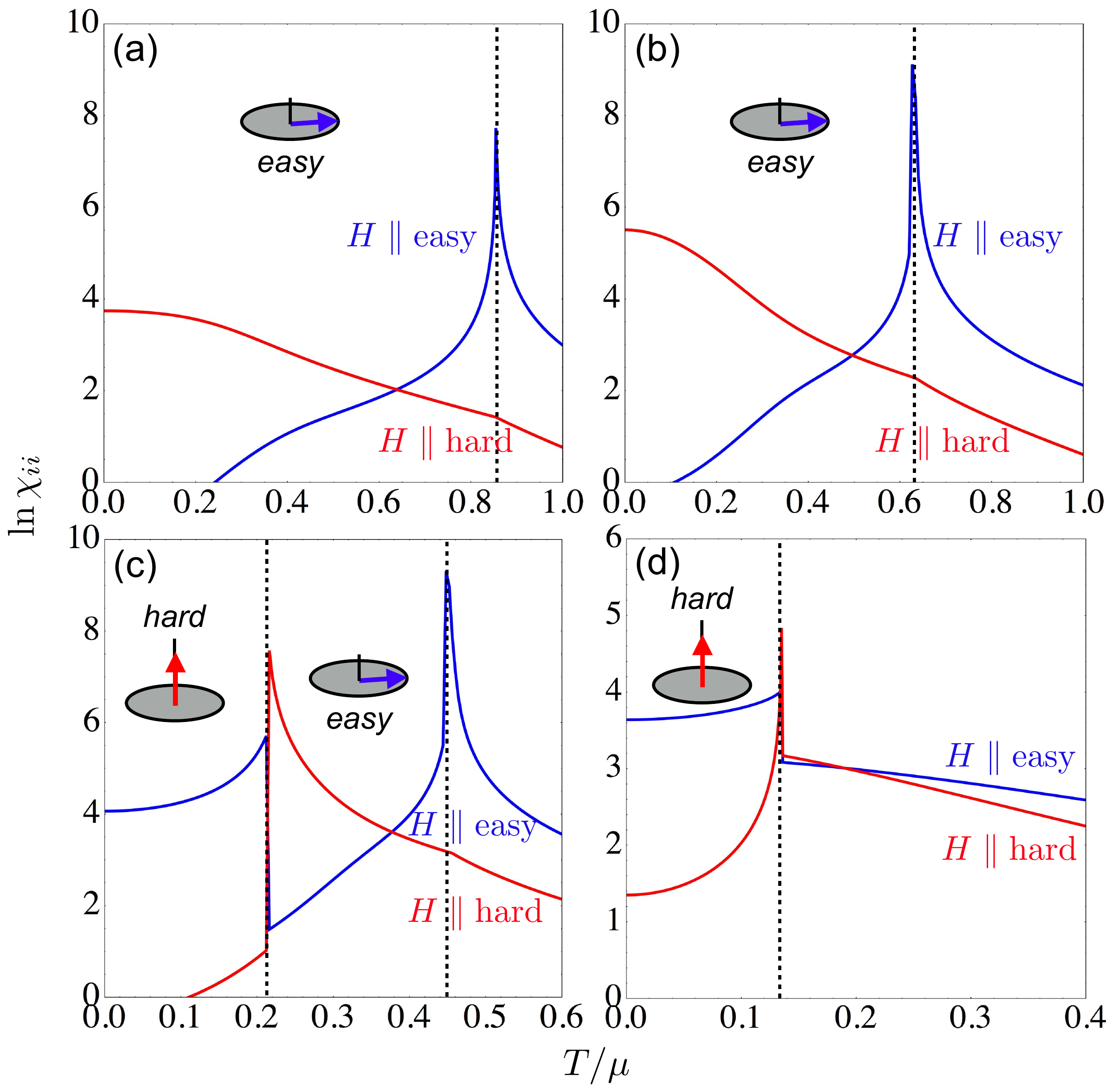}
\caption{(color online). Logarithm of the magnetic susceptibilities $\chi_{ii}=\partial m_i/\partial h_i$ for fields in the easy plane (blue) and along the hard axis (red) as 
a function of temperature for decreasing values of the electron repulsion: $1/(\rho U) = $ 0.093 (a), 0.0945 (b), 0.098 (c), 0.103 (d).}
\label{fig.3}
\end{center}
\end{figure}

Note that the region of FM order along the hard axis shrinks with increasing band splitting $\Delta$ and is completely suppressed 
beyond a critical value $\Delta_c$ (see inset of Fig.~\ref{fig.2}). Since the competing directional dependent terms $F_\textrm{mf}^{(\lambda)}$ and $F_\textrm{fl}^{(\lambda)}$
are of the same order in the Hund interaction $J_H$ and SO coupling $\lambda$, these parameters only have a small effect on the stability of the hard-axis FM.

Depending upon the value of $U$, which can be viewed as a proxy for tuning parameters like pressure or chemical doping, 
there are three possible scenarios of phase transitions as a function of decreasing temperature. 

(i) For sufficiently large $U$, we expect a single transition from a paramagnet to an easy plane FM that is stable down to $T=0$. Depending on the separation 
of $P$ and $Q$ this transition could be continuous or first-order.

(ii) Over an intermediate range of $U$ we expect a sequence of two transitions, first from a paramagnet into an easy-plane FM (1st or 2nd order) and 
then at lower temperatures into a FM state with moments along the hard direction. The transition at which the moment direction 
switches is always discontinuous. 

(iii) Decreasing  $U$ further, we find a single first-order transition from a paramagnet into a hard-axis FM. 

To make contact with experiments, we calculate the magnetic susceptibilities $\chi_\perp$ and $\chi_\parallel$
for fields along the easy and hard directions. In Fig.~\ref{fig.3} the evolution of the susceptibilities with temperature
is shown for different values of the electron repulsion $U$. Panels (a) and (b) correspond to case (i) with single continuous transitions into the
easy plane FM. At the transition, the easy plane susceptibility $\chi_\perp$ diverges. The proximity of the system to a fluctuation-driven reorientation of the 
moments is apparent from the increase in the hard axis susceptibility as the temperature is reduced.  Reducing $U$, this increase 
becomes more pronounced. Case (ii)  is shown in panel (c). At the first-order transition between the two FM
states we find an inversion of $\chi_\perp$ and $\chi_\parallel$, characteristic of the switching of the moment direction. 

Finally, Fig.~\ref{fig.3}(d) shows the temperature dependence of the susceptibilities in the regime (iii) where the system exhibits a direct transition from a paramagnet to 
a FM state with moments along the hard direction, exactly as it has been observed experimentally \cite{Lausberg+13,Steppke+13}. We find a similar 
characteristic crossing of susceptibilities at a temperature $T^*$ slightly above the ordering temperature $T_c$. As we increase $U$ and move closer towards the point 
$Q$ (see Fig.~\ref{fig.2}), the temperatures $T^*$ and $T_c$ merge.

{\bf Discussion and Conclusions.} We have described a generic mechanism whereby transverse quantum fluctuations drive unexpected magnetic behavior in itinerant FMs with 
magnetic anisotropy. At low temperatures, these systems can magnetize along directions that are unfavorable at higher temperatures.  
As a proof of principle we have explicitly demonstrated this for a simple two-band model with an 
easy-plane anisotropy generated by SO coupling.  We found that close to the quantum critical point the moments switch to the magnetic hard axis in order to maximize the phase space for transverse spin fluctuations -- a `quantum indian rope trick'! \cite{indian_rope}. 
While we treat the SO perturbatively, the enhancement of hard-axis ordering with increasing SO coupling suggest that our results also apply to systems with strong SO
interaction.

Related phase transitions are also possible. Quantum fluctuations could also stabilize a modulated spiral state which preempts the first-order transition into the homogeneous 
FM \cite{Conduit+09,Karahasanovic+12}. Since planar spirals are compatible with magnetic easy-plane anisotropies, one might find 
systems that show spiral order which then becomes unstable toward a hard-axis FM at even lower temperatures. 
For a tetragonal system with magnetic easy-axis anisotropy we expect a similar moment reorientation as for the easy-plane case studied in this Letter. Simply because by flipping the moments into the hard plane, the easy direction becomes available for transverse spin fluctuations. Orthorhombic systems with three inequivalent magnetic directions might offer even more interesting phase behavior with a two-step moment
reorientation. 

With recent  experimental advances it is now possible to study the phase reconstruction near quantum critical points down to extremely low temperatures. 
There are several materials in which FM ordering along a hard magnetic direction has been discovered. The most promising is $\text{Yb} \text{Ni}_4 \text{P}_2$ 
\cite{Steppke+13} because of its close proximity to a quantum phase transition. The observed crossing of susceptibilities  
slightly above the ordering temperature is in qualitative agreement with our prediction based on a simple itinerant model. However, the measured critical exponents 
are inconsistent with a pure itinerant model \cite{Steppke+13}. In fact, $\text{Yb} \text{Ni}_4 \text{P}_2$ exhibits quasi one-dimensional spin 
chains, which couple to the conduction electrons, and is a Kondo-lattice system with strong interactions between conduction- and localized $f$-electrons. 
While the non-analytic free energy corrections are expected to have similar effects in systems with local 
moments \cite{Kirkpatrick+12}, a quantitative analysis based on a realistic microscopic model for $\text{Yb} \text{Ni}_4 \text{P}_2$ would be desirable.

We point out that the suggested mechanism for the moment switching close to FM quantum critical points does not require frustration. It is expected to be generic since 
it stems from the same non-analytic free energy corrections that are responsible for fluctuation-induced first-order behavior in practically all itinerant FMs, irrespective of
microscopic details.  It is possible to go beyond our simple model to include frustration, the coupling to local moments, and potentially even Kondo physics. The interplay of 
these extra ingredients with electronic low energy-fluctuations might yield a plethora of novel and exciting ordering phenomena.

{\bf{Acknowledgements}}
The authors benefitted from stimulating discussions with M. Brando, A. Steppke, and M. Vojta. This work has been supported by the EPSRC through grant EP/I004831/2.


\begin{thebibliography}{32}
\bibitem[{der()}]{derivativetrick}
\bibinfo{note}{To compute the $\Gamma^{(1)}_{\nu_1,\nu_2, \nu_3}$ term, we
  write it as \begin{eqnarray} \Gamma^{(1)}_{\nu_1\nu_2\nu_3} & = &
  \frac{D^{(1)}_{\nu_1\nu_2}\Pi^{(1)}_{\nu_1\nu_2}}{2 \zeta_{\nu_2 \nu_3}} -
  \frac{\left(\Pi^{(1)}_{\nu_1\nu_2} - \tilde{\Pi}_{\nu_1\nu_2}
  \right)}{\zeta_{\nu_2 \nu_3}^2} \nn \\ \zeta_{\nu_2 \nu_3} & = &
  \Delta+U(\nu_2\mt_1-\nu_3\mt_2)\nn, \end{eqnarray} where the derivative
  operators $D^{(1)}_{\nu\nu}=\partial_{\mu_1}$ and
  $D^{(1)}_{\nu\overline{\nu}}=\partial_{\mu_1}-\frac{\nu}{U}\partial_{\mt_1}$
  and $\tilde{\Pi}_{\nu_1\nu_2}(q) = T\sum_{\omega}\int_\bk
  G_{1,\nu_1}(k+q)G_{2,\nu_2}(k)$. Substituting into the fluctuation term gives
  \begin{equation} F_\textrm{fl}^{(\lambda)} \approx -\frac{\lambda^2 U^2
  T}{4\Delta^2}\left(\mathcal{P}_\parallel-\mathcal{P}_\perp \right) \mt_2
  \partial_{\mt_1} \sum_{\tilde{\omega}}\int_\bq \Pi^{(1)}_{+-}\Pi^{(1)}_{-+}
  \nn \end{equation} which shows that the dominant terms originate from
  transverse spin fluctuations and leads to the result given in the text.}
\bibitem[{ind()}]{indian_rope}
\bibinfo{note}{In the extreme version of the Indian-RopeTrick a rope rises from
  a basket and a helper of the Fakir then climbs up the rope. Finally, the
  helper would descend or even disappear from view. This is either a hoax or
  true magic. It is however possible to perform an Indian-Rope Trick with a
  chain of inverted pendula, stabilized by vibrations of the base point
  \cite{Acheson93,Acheson+93}.}
\end{thebibliography}
\end{document}